\documentclass[mathleft
]{an}
\usepackage{graphicx}
\usepackage{times}
\def\asca{{\it ASCA}}

\def\xmm{{\it XMM-Newton}}
\begin{document}

\Pagespan{789}{}
\Yearpublication{2006}%
\Yearsubmission{2005}%
\Month{11}%
\Volume{999}%
\Issue{88}%

\title{An XMM-Newton survey of broad iron lines in AGN}

\author{K. Nandra\inst{1}\fnmsep\thanks{Corresponding author:
  \email{k.nandra@imperial.ac.uk}\newline}
\and
P.M. O'Neill\inst{1}
\and
I.M. George\inst{2,3}
\and
J.N. Reeves\inst{3,4}
\and  
T.J. Turner\inst{2,3}
}
\titlerunning{Broad lines in AGN}
\authorrunning{K. Nandra et al.}
\institute{
Astrophysics Group, Blackett Laboratory, Imperial College London, London SW7 2AZ, UK
\and 
Department of Physics and Astronomy, University of Maryland, Baltimore County, Hilltop Circle, Baltimore, MD, USA
\and 
Laboratory for High Energy Astrophysics, NASA/Goddard Space Flight Center, Code 660, Greenbelt, MD 20771, USA
\and
Department of Physics \& Astronomy, Johns Hopkins University, Baltimore, MD 
}

\received{XXXXX}
\accepted{XXXXX}
\publonline{later}


\abstract{%
We report on the iron K$\alpha$ line properties of a sample of Seyfert galaxies observed with the {\it XMM-Newton} EPIC pn instrument. Using a systematic and uniform analysis, we find that complexity at iron-K is extremely common in the \xmm\ spectra. Once appropriate soft X-ray absorption, narrow 6.4~keV emission and associated Compton reflection are accounted for,  $\sim 75$~\% of the sample show an improvement when a further Gaussian component is introduced. The typical properties of the broad emission are both qualitatively and quantitatively consistent with previous results from \asca. The complexity is in general very well described by relativistic accretion disk models. In most cases the characteristic emission radius is constrained to be within $\sim 50$~$R_{\rm g}$, where strong gravitational effects become important.  We find in about 1/3 of the sample the accretion disk interpretation is strongly favoured over competing models. In a few objects no broad line is apparent. We find evidence for emission within $6 R_{\rm g}$ in only two cases, both of which exhibit highly complex absorption. Evidence for black hole spin based on the X-ray spectra therefore remains tentative.  
 }
\maketitle

\section{Introduction}

Observations with \asca\ showed complex, broad emission from iron to be very common in Seyfert galaxies (Nandra et al. 1997). These lines can be interpreted as emission from a relativistic accretion disk, in which case they represent a powerful probe of the  strong gravity regime around black holes (Fabian et al. 1989; Stella 1990). The most celebrated case is MCG-6-30-15, where the broad, skewed line seen with \asca\ is of high signal-to-noise ratio, and the disk line interpretation is apparently robust (Tanaka et al. 1995; Fabian et al. 1995). Several other high quality profiles from {\it ASCA} also showed broad, relativistic lines (e.g. George et al. 1998; Nandra et al. 1999; Done et al. 2000). 

Since the launch of \xmm, it has been possible to obtain even higher quality data on these broad emission lines. Early results confirmed relativistic emission in some cases, including MCG-6-30-15 (e.g. Wilms et a. 2001; Fabian et al. 2002; Vaughan \& Fabian 2004), but in others no broad line was detected (e.g. Gondoin et al. 2001; Pounds et al. 2003; Bianchi et al. 2004). In yet others, complexity has been observed around iron-$K$, but the interpretation as relativistic disk emission has been challenged. One specific suggestion is that absorption by a high column, high ionization warm absorber can mimic the ``red wing'' characteristic of an accretion disk line (Reeves et al. 2004). 

The absence of a comprehensive and systematic survey of the X-ray spectra of Seyferts observed by \xmm\ prevents firm conclusions being drawn as to the prevalence of broad iron lines in AGN and the robustness of their interpretation. Here we present preliminary results from such a study, the full results of which will be presented in a forthcoming paper (Nandra et al. 2006, in preparation). 

\section{The sample and data analysis}

Our sample is culled from pointed AGN observations in the \xmm\ archive.  We examine only local AGN($z<0.05$) and exclude Seyfert 2 galaxies and radio loud objects. Furthermore we choose only the objects with the highest number of counts in the 2-10 keV band, to maximise the signal-to-noise ratio around the iron line. The sample reported here consists of 41 observations of 30 objects.

An important feature of our work is that we have performed a well-defined, uniform analysis, with conservative selection cirteria and using the latest available calibrations. The techniques are fully decribed in O'Neill et al. (2006, in preparation), but compared to much of the previous work the improvements include: a) consistent definition of source and background regions for each observation b) well defined and conservative background rejection c) precise definition of good-time intervals d) standardised spectral grouping related to the instrumental resolution. We restrict our analysis to the pn instrument. For observations with significant pileup we use only the pattern 0 events. Spectral fits are undertaken in the 2.5-10 keV range only, to minimize complications due to absorption and soft excess emission, and avoid the instrumental calibration feature around 2.2 keV. 

\begin{table}
\caption{Comparison between mean parameters for broad lines determined by \asca\ (Nandra et al. 1997) and \xmm\ (this work). Note that that \asca\ fits did not account for a distant narrow component of the Fe K$\alpha$, nor did they include a warm absorber. The fraction of objects in which the F-test indicates a 99\% improvement is given, along with the mean Energy, Gaussian $\sigma$ and equivalent width. 
\label{tab:mpars}}
\begin{tabular}{lllll}\hline
 & Fraction & Energy & Width & EW \\
 & & (keV) & (keV) & (eV) \\
  & (1) & (2) & (3) & (4) \\ 
\hline
{\it ASCA} 	& 77\% & $6.34\pm 0.04$ & $0.43 \pm 0.12$ &  $160 \pm 30$ \\
{\it XMM}           & 73\% & $6.32\pm 0.05$ & $0.36\pm 0.04$  & $108\pm 12$ \\
\hline
\end{tabular}
\end{table}
\section{Results}

\subsection{Base model}

While we have excluded the most heavily obscured objects (Seyfert 2s) from our sample, there remains a possibility that absorption can have a significant effect even on the spectra above 2.5 keV. We account for this by fitting an XSTAR (Kallman et al. 2004) photoionization model to the spectra,excluding the iron band (4.5-7.5 keV).  Where the the fit improves significantly at 95\% confidence, according to the F-test, this XSTAR component is included in all subsequent fits, with free $N_{\rm H}$ and ionization parameter. It is now also known that many AGN exhibit narrow cores to their iron K$\alpha$ lines (e.g. Yaqoob \& Padhmanhaban 2004). These are thought to arise from very distant material such as the torus (e.g. Krolik \& Kallman 1987; Awaki et al. 1991). If so they will be accompanied by continuum Compton reflection. We therefore include in all fits below a neutral Compton reflection component appropriate for a slab geometry (Magdziarz \& Zdziarski 1995), with accompanying Fe K$\alpha$, Fe K$\beta$ and Ni K$\alpha$, line emission (George \& Fabian 1991) and a Compton shoulder (Matt 2002). The emission lines and reflection are all incorporated in a single model with solar abundances. We assume an inclination of $60^{\circ}$ for the slab and hence the reflection is characterized by a single parameter, $R = \Omega/2\pi$, where $\Omega$ is the solid angle subtended by the slab at the illuminating source. 

\begin{figure}
\includegraphics[angle=270,width=80mm]{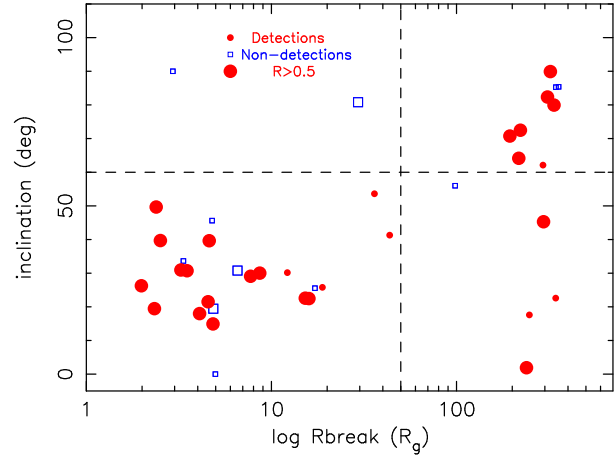}
\caption{Characteristic emission radius for the relativistic iron K$\alpha$ lines versus disk inclination. 
}
\label{fig:rbreak}
\end{figure}

\subsection{Simple parameterization of the broad emission}

To provide a simple, model-independent characterization of further complexity in the iron band, we have added a broad Gaussian to the fits described above. A significant improvement to the fit was found in 30 of the 41 observations, and 22 of the 30 objects. Clearly, complexity at iron K$\alpha$ is extremely common in Seyferts. A comparison between the mean parameters of the broad Gaussian fits to the \asca\ data (Nandra et al. 1997) and our new \xmm\ sample, is given in Table~\ref{tab:mpars}. There is remarkable agreement in all cases, with the exception that the line equivalent widths in the \asca\ sample are about $50$~\% higher. This difference can be attributed to the fact that the narrow line cores were deconvolved in the \xmm\ fits, but not with \asca. 

The energies of the broad lines seen with \xmm\ clearly indicate that they are associated with iron, as they are very close to the expected energy, but there is some evidence that the typical energy is redshifted compared to the neutral value. The lines are usually quite broad, with $<\sigma>=0.36$~keV or 40,000 km s$^{-1}$ FWHM. Significant dispersion is seen in all the measured quantities, however, which confirms the result from \asca\ that there is a wide variety of line profiles, and takes this further in that the variation from object-to-object cannot be attributed solely to varying relative contributions of the narrow core and broader emission. It should also be noted that in 5 of the fits, the width of the ``broad" gaussian component is $<10,000$ km s$^{-1}$. These lines could plausibly arise from the optical broad line region (BLR), rather than the inner disk. 

\subsection{Disk line models}

We have tested explicitly whether the complex line shapes seen in the \xmm\ spectra can be accounted for with a relativistic accretion disk. We do this by adding an additional, neutral reflection component with Fe and Ni line emission as above, but this time apply relativistic blurring (Laor 1991; Fabian et al. 2002).  
Rather than leave all the parameters free, we initially chose to fix the inner and outer radii at $R_{\rm i}=6 R_{\rm g}$ and $R_{\rm o}=400 R_{\rm g}$. We adopt an emissivity law appropriate for a point source above a slab in a Newtonian geometry, which can be approximated as a broken power law. The adopted emissivity depends on $R^{-q}$, with $q=0$ within and $q=3$ outside some break radius $R_{\rm br}$. This represents the  characteristic radius where the majority of the line emission originates, so can be used to assess whether relativistic effects are important. The inclination and reflection fraction are left as free parameters too. The relativistically blurred model improves the fits significantly in $\sim 75$\% of the observations and indeed gives markedly better fits than a Gaussian in several cases. 

The characteristic emission radius ($R_{\rm br}$) is plotted against the inclination in Fig.~\ref{fig:rbreak}. The bottom left part of this diagram is where we expect ``classic'' disk lines to occur. Here the emission is concentrated in the innermost regions ($<50$~R$_{\rm g}$) and the inclination is relatively low, such that much of the emission is redshifted. The upper left portion is where we expect weak and very broad lines from highly inclined disks. It is sparsely populated, which is expected as such lines are difficult to detect.  The upper right portion shows several strong disk lines with apparently high inclinations but at relatively large radii. This indicates that the lines are broad but predominantly towards the blue rather than the red. These are likely candidates for a highly ionized disk, which is in reality at lower inclination than inferred in fits which assume the disk is neutral. Finally, at the bottom right of the diagram we see emission at low inclination and large radius. In these objects the lines will be relatively narrow and not strongly shifted. For these, there is no requirement for the line to arise in the inner accretion disk and they may come from more distant material, such as the optical BLR.

Using this model, we can also assess the evidence for black hole spin. A simple test is to repeat  the fits using an inner radius of $1.235 R_{\rm g}$, appropriate for a Kerr Black hole with $a/M=0.998$, as opposed to the Schwarzschild value of $6 R_{\rm g}$. Only two of the spectra showed an appreciable improvement with such a model. In both, NGC 3783 and NGC 4151, there is complex absorption which strongly affects the spectrum above 2.5 keV (Reeves et al. 2004; Schurch et al. 2004). We therefore consider the evidence for maximal Kerr black holes to be tentative, leaving it an open question as to whether black holes in AGN are generally rotating.

\subsection{Comparison with alternative models}

Some recent studies have suggested alternatives to the relativistic disk model for broad iron lines in AGN. A number of objects show no evidence for broad emission at all, including some in our sample. In others, it may be possible to model the ``red wing'' as a high ionization warm absorber (Reeves et al. 2004), and the ``blue wing'' with blends of narrow lines. To test this, we have fitted a model comprising a high ionization warm absorber (in  {\it addition} lower ionization gas already included), with three narrow emission lines, two fixed at the energies appropriate to helium and  hydrogen--like iron and another intermediate (6.4-6.7 keV) line with free energy. Once again neutral, unblurred reflection is also included to account for any narrow emission at 6.4 keV. 

 \begin{figure}
\vspace{-0.4in}
\includegraphics[angle=0,width=80mm]{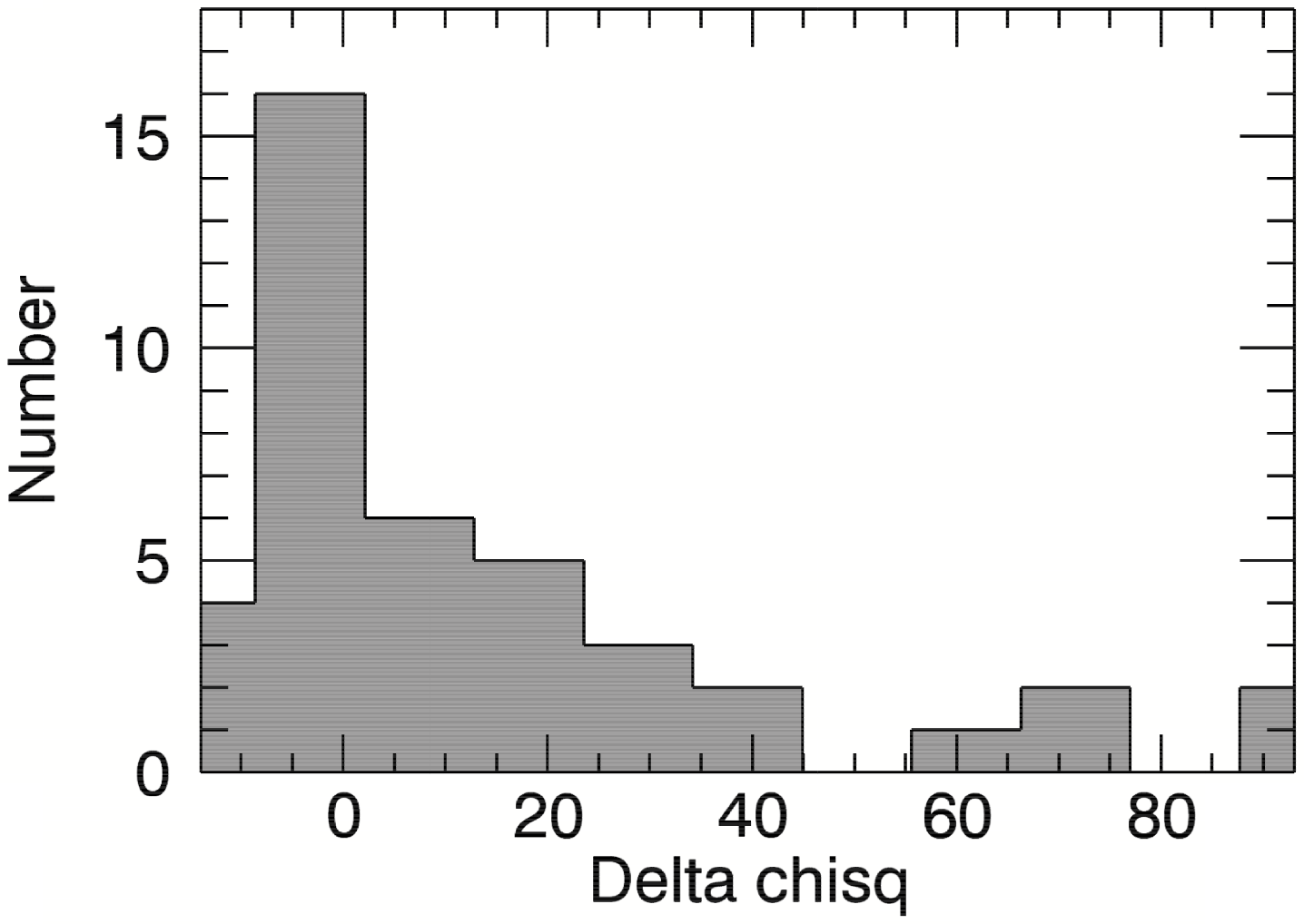}
\vspace{-0.4in}
\caption{Difference in $\chi^{2}$ between the relativistic disk line model and an alternative model comprising a high ionization warm absorber, and a blend of narrow lines. All fits include both line and continuum from a distant neutral reflector and a soft X-ray warm absorber where needed.  The disk line model has 3 fewer free parameters than the alternative, but provides a dramatically better fit in a large number of cases (see Fig.~\ref{fig:dream})}
\label{fig:dchi}
\end{figure}

A comparison of the $\chi^{2}$ values is given in Fig.~\ref{fig:dchi}. The alternative model has 3 more free parameters than the disk line model, but provides a substantially worse fit in a large number of objects. In a few cases the alternative model fits a little better, but not substantially so considering the larger number of free parameters. 

\begin{figure*}
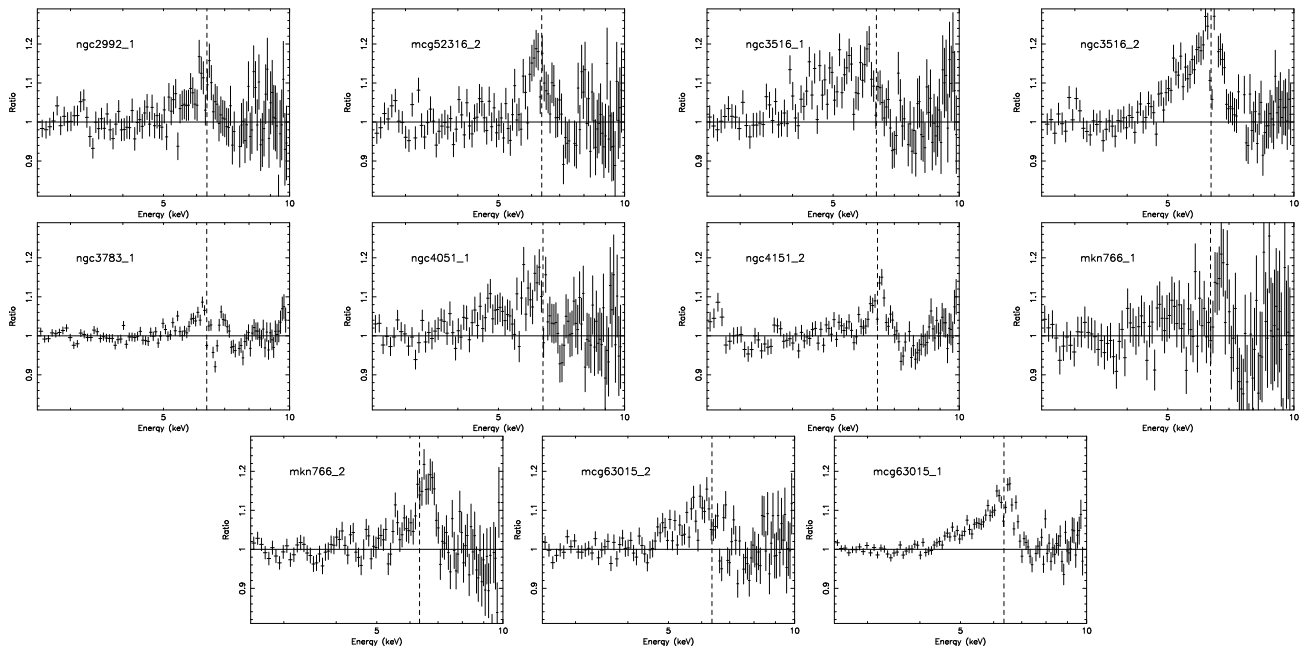

{
\includegraphics[angle=-90,width=38mm]{ngc2992_1_bl_plot.ps}
\includegraphics[angle=-90,width=38mm]{mcg52316_2_bl_plot.ps}
\includegraphics[angle=-90,width=38mm]{ngc3516_1_bl_plot.ps}
\includegraphics[angle=-90,width=38mm]{ngc3516_2_bl_plot.ps}
\includegraphics[angle=-90,width=38mm]{ngc3783_1_bl_plot.ps}
\includegraphics[angle=-90,width=38mm]{ngc4051_1_bl_plot.ps}
\includegraphics[angle=-90,width=38mm]{ngc4151_3_bl_plot.ps}
\includegraphics[angle=-90,width=38mm]{mkn766_1_bl_plot.ps}
\centerline{
\includegraphics[angle=-90,width=38mm]{mkn766_2_bl_plot.ps}
\includegraphics[angle=-90,width=38mm]{mcg63015_1_bl_plot.ps}
\includegraphics[angle=-90,width=38mm]{mcg63015_2_bl_plot.ps}}
}
\caption{The relativistic line ``dream team":  data/model ratios are shown in cases where the broad line emission is constrained to come from within $R_{\rm br} <20 R_{\rm g}$, and for which the alternative model narrow line blends plus warm absorber gives a worse fit by $\Delta\chi^{2}>10$. The reference continuum is a power law, with a warm absorber where necessary. Narrow line emission at 6.4 keV, with associated Compton reflection, is also included so the above ratio plots may underestimate any contribution at 6.4 keV from the accretion disk line. 
 \label{fig:dream}}
\end{figure*}

From the consideration of this alternative model, and the results from Fig~\ref{fig:rbreak}, we can define a sample of robust relativistic lines for which disk models both indicate a small characteristic radius, and fit much better than the alternative. There are 11 spectra of 9 objects satisfying the criteria that $R_{br}<20$~$R_{\rm g}$ and $\Delta\chi^{2}>10$ for the relativistic model compared to the alternative model (despite having 3 {\it fewer} free parameters). The line profiles are shown in Fig.~\ref{fig:dream}.

\section{Discussion and conclusions}

Our systematic \xmm\ survey should serve to clear up some of the controversy about how often broad emission lines from an accretion disk can be claimed robustly in AGN. For Seyferts, at least, complexity at iron K$\alpha$ is seen in about 3/4 of objects and this complexity is always interpretable in terms of an accretion disk model. In about 1/3 of our sample, that interpretation is clearly preferred over competing models. In a few cases with high signal-to-noise ratio the relativistic emission appears to be absent, but 
great caution needs to be exercised before this can be concluded definitively, as even with \xmm\ very good statistics are required (Guainazzi et al., this volume). In cases where broad emission appears to be absent, the disk may simply be highly inclined, such that the line is very broad and weak. Alternatively, the inner disk may be hot and/or highly ionized (Nayakshin 2000), which can also account for cases where broad emission is present, but indicative of a relative large characteristic radius ($\sim 100$~$R{\rm g}$). Alternatively, the lack of broad emission seen in a given observation may be due to line profile variability (Longinotti et al. 2004).

Perhaps surprisingly, we have not yet found any strong evidence for black hole spin. This contrasts with some previous studies indicating maximally rotating holes in MCG-6-30-15 (Wilms et al. 2001) and some black hole binaries (e.g. Miller et al. 2002).  This is probably due to our conservative approach in consideration of distant reflection and complex absorption. On the other hand, our observations provide no evidence {\it against} rapidly rotating black holes in AGN and while it has been pointed out in several previous studies that complex absorption can mimic very broad lines (e.g. Done \& Gierlinski 2006), it is important to bear in mind that the converse is also true.

Our main conclusion, however, is that the accretion disk interpretation for broad iron K$\alpha$ lines in AGN appears to be robust. The implication is that the potential for X-ray observations, particularly with {\it XEUS} and {\it Con--X}, to reveal new information about the innermost regions of accreting black holes may well be realised.

\acknowledgements
We thank  Tim Kallman for help with XSTAR; PPARC and the Leverhulme Trust for financial support and gratefully acknowledge those who built and operate the \xmm\ satellite.


\end{document}